\documentclass[letter,bibyear]{aa} 

\usepackage{epsfig}
\usepackage{amsmath}
\usepackage{subfigure}
\usepackage{natbib}
\usepackage{multirow}
\usepackage{color}
\usepackage{longtable}
\usepackage{graphicx}
\usepackage{epstopdf}
\usepackage{booktabs}
\usepackage{txfonts}

\newcommand{\jms}{J.~Mol.~Spectrosc.}   
\newcommand{\jmst}{J.~Mol.~Struct.}   

\newcommand{\kms}{km s$^{-1}$}

\newcommand{\once}{10$^{11}$\,cm$^{-2}$}
\newcommand{\doce}{10$^{12}$\,cm$^{-2}$}

\begin{document}

\title{The sulphur saga in TMC-1: Discovery of HCSCN and HCSCCH
\thanks{Based on observations carried out
with the Yebes 40m telescope (projects 19A003,
20A014, 20D023, and 21A011). The 40m
radio telescope at Yebes Observatory is operated by the Spanish Geographic
Institute,
(IGN, Ministerio de Transportes, Movilidad y Agenda Urbana).}}
\author{
J.~Cernicharo\inst{1},
C.~Cabezas\inst{1},
Y.~Endo\inst{2},
M.~Ag\'undez\inst{1},
B.~Tercero\inst{3,4},
J.~R.~Pardo\inst{1},
N.~Marcelino\inst{1},
P.~de~Vicente\inst{3}
}

\institute{Grupo de Astrof\'isica Molecular, Instituto de F\'isica Fundamental (IFF-CSIC),
C/ Serrano 121, 28006 Madrid, Spain\\ \email jose.cernicharo@csic.es,carlos.cabezas@csic.es
\and Department of Applied Chemistry, Science Building II, National Yang Ming Chiao Tung University, 1001 Ta-Hsueh Rd., Hsinchu 30010, Taiwan
\and Centro de Desarrollos Tecnol\'ogicos, Observatorio de Yebes (IGN), 19141 Yebes, Guadalajara, Spain
\and Observatorio Astron\'omico Nacional (IGN), C/Alfonso XII, 3, 28014, Madrid, Spain
}
\date{Received; accepted}

\abstract{We report the detection, 
for the first time in space, of cyano thioformaldehyde (HCSCN) 
and
propynethial (HCSCCH) towards the starless core TMC-1. Cyano thioformaldehyde presents a series 
of prominent $a$- and $b$-type lines,
which are the strongest previously unassigned features in our Q-band line survey of TMC-1. Remarkably, 
HCSCN is four times more abundant
than cyano formaldehyde (HCOCN). On the other hand, HCSCCH is five times less abundant than
propynal (HCOCCH). Surprisingly, we find an abundance ratio HCSCCH/HCSCN of $\sim$\,0.25, in 
contrast with most other ethynyl-cyanide pairs of molecules for which the CCH-bearing species 
is more abundant than the CN-bearing one. We discuss the formation of these molecules in terms 
of neutral-neutral reactions of S atoms with CH$_2$CCH and CH$_2$CN radicals as well as of CCH and CN 
radicals with H$_2$CS. The calculated abundances for the sulphur-bearing species are, however, 
significantly below the observed values, which points to an underestimation of the abundance 
of atomic sulphur in the model or 
to missing formation reactions, such as ion-neutral reactions.}


\titlerunning{The sulphur saga in TMC-1}
\authorrunning{Cernicharo et al.}

\maketitle

\section{Introduction}

The chemistry of sulphur-bearing molecules in cold interstellar clouds is an active area of 
research \citep{Vidal2017,Cernicharo2021a,Cernicharo2021b}. Recently, \citet{Cernicharo2021b} 
reported the detection
of five new sulphur-bearing species in TMC-1, a source
that presents an interesting carbon-rich chemistry that leads to the formation of long neutral carbon 
chain
radicals and their anions, cyanopolyynes (see Cernicharo et al. 2020a,b and references therein), 
pure hydrocarbon cycles, and polycyclic aromatic hydrocarbons \citep{Cernicharo2021c,Burkhardt2021,McCarthy2021}. A 
large variety of S-bearing molecules are also present in this source, which leads to new 
questions regarding the formation routes to these species. It is clear that many ion-neutral 
and neutral-neutral reactions involving S-bearing species have to be studied to improve 
the reliability of chemical models \citep{Petrie1996,Bulut2021}. Unveiling new 
sulphur-bearing species can also help to understand the chemistry of sulphur by comparing 
the observed abundances with those predicted by chemical models.

\begin{figure}[]
\centering
\includegraphics[scale=0.3,angle=0]{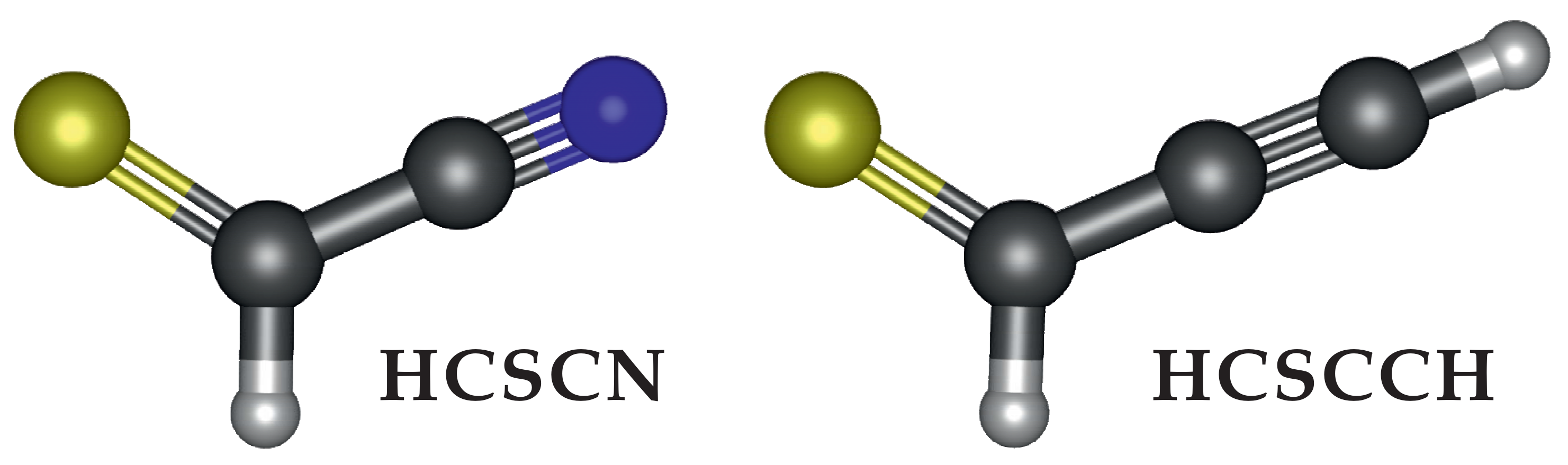}
\caption{Structure of cyano thioformaldehyde, HCSCN, and
ethynyl thioformaldehyde, HCSCCH, also known as propynethial.
}
\label{fig_structures}
\end{figure}

In this letter we report the discovery of two new sulphur-bearing molecules: HCSCN (cyano 
thioformaldehyde)
and HCSCCH (propynethial). Figure~\ref{fig_structures} shows the structure of these molecules.
The derived abundances are compared with the oxygen analogues of these species,
HCOCN (cyano formaldehyde) and \mbox{HCOCCH} (propynal). We discuss plausible reactions that could lead to the formation of these
species with the aid of a chemical model.

\begin{figure*}[]
\centering
\includegraphics[scale=0.55,angle=0]{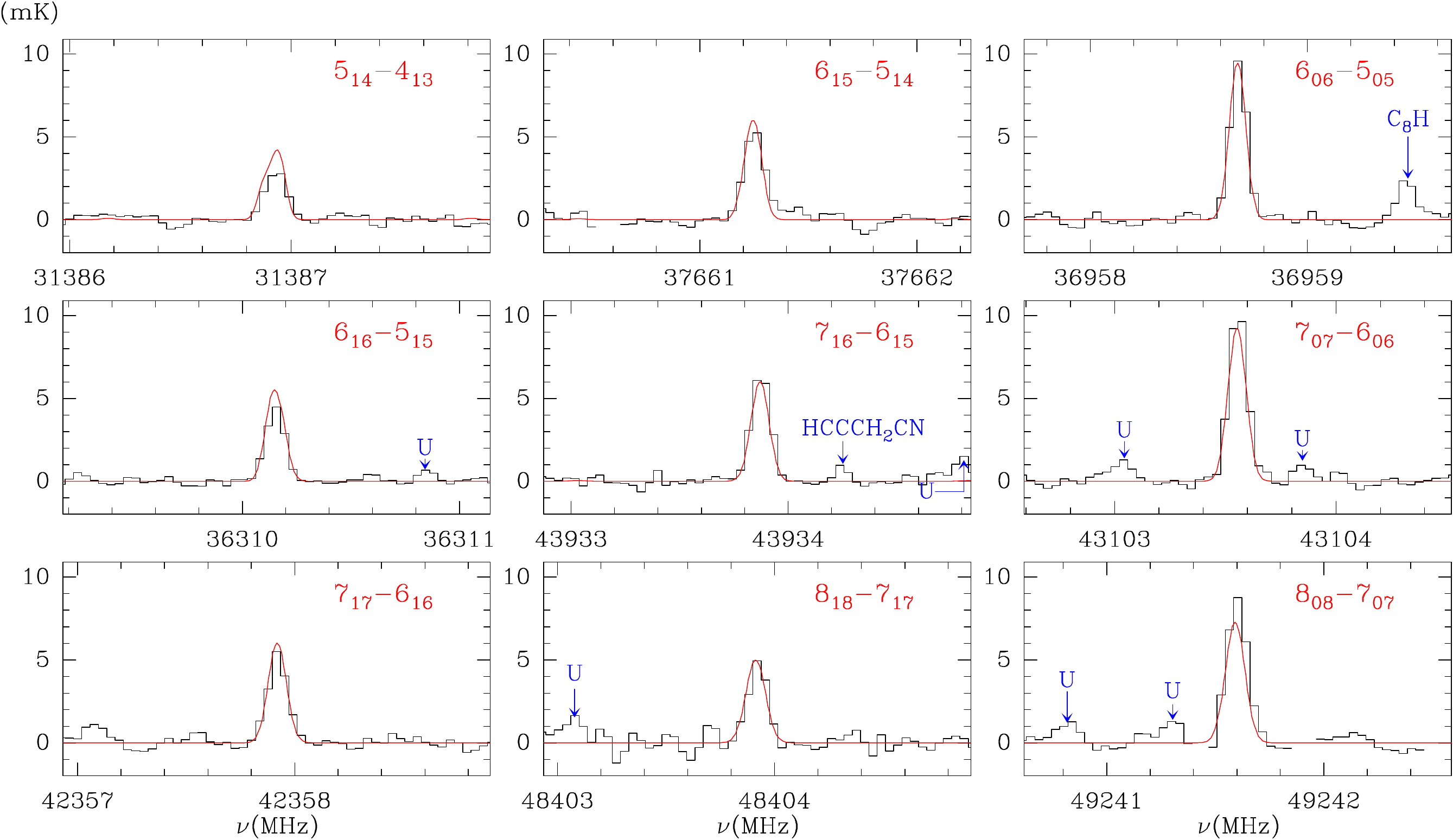}
\caption{Selected $a$-type transitions of HCSCN.
The abscissa corresponds to the rest frequency of the lines assuming a
local standard of rest velocity of the source of 5.83 km s$^{-1}$. Frequencies and intensities for the observed lines
are given in Table \ref{obs_line_parameters}.
The ordinate is the antenna temperature, corrected for atmospheric and telescope losses, in millikelvin.
The quantum numbers for each transition are indicated
in the upper left corner of the corresponding panel.
The red lines show the computed synthetic spectrum for this species for T$_r$=5 K and
a column density of 1.3$\times$\doce. Blanked channels correspond to negative features produced
in the folding of the frequency switching data.}
\label{fig_type_a}
\end{figure*}

\section{Observations}
\label{observations}
New receivers, built within the Nanocosmos project\footnote{\texttt{https://nanocosmos.iff.csic.es/}}
and installed at the Yebes 40m radio telescope, were used
for the observations of TMC-1. A detailed description of the system is given by \citet{Tercero2021}.
The Q-band receiver consists of two high electron mobility transistor cold amplifiers covering the
31.0-50.3 GHz band with horizontal and vertical polarisations.
The backends are $2\times8\times2.5$ GHz fast Fourier transform (FT) spectrometers
with a spectral resolution of 38.15 kHz
that provide the whole coverage of the Q-band in both polarisations. The main beam efficiency varies from 0.6 at
32 GHz to 0.43 at 50 GHz.

The line survey of TMC-1 ($\alpha_{J2000}=4^{\rm h} 41^{\rm  m} 41.9^{\rm s}$ and
$\delta_{J2000}=+25^\circ 41' 27.0''$)
in the Q-band was performed in several sessions. The first results
\citep{Cernicharo2020a,Cernicharo2020b} were based on two observing runs performed in
November 2019 and February 2020. Two different frequency coverages were achieved,
31.08-49.52 GHz and 31.98-50.42 GHz, in order to check that no
spurious spectral ghosts are produced in the down-conversion chain.
Additional data were taken in October 2020, December 2020, and January-April 2021.
The observing procedure was frequency switching with a frequency throw of 10\,MHz for the two first runs and of
8\,MHz for all the others.
Receiver temperatures in the runs achieved in 2020 vary from 22 K at 32 GHz
to 42 K at 50 GHz. Some power adaptation in the down-conversion chains reduced
the receiver temperatures in 2021 to 16\,K at 32 GHz and 25\,K at 50 GHz.
The sensitivity of the survey varies along the
Q-band between 0.25 and 1 mK, which is a considerable improvement over previous line surveys in the 31-50 GHz frequency range
\citep{Kaifu2004}. The acronym we have assigned to this line survey is
QUIJOTE (Q-band Ultrasensitive Inspection Journey to the Obscure Tmc-1 Environment).

The intensity scale used in this work, the antenna temperature
($T_A^*$), was calibrated using two absorbers at different temperatures and the
atmospheric transmission model ATM \citep{Cernicharo1985, Pardo2001}.
Calibration uncertainties of 10~\% were adopted.
All data were analysed using the GILDAS package\footnote{\texttt{http://www.iram.fr/IRAMFR/GILDAS}}.

\section{Assignment of unidentified lines to HCSCN}
\label{assig_1st_step}

Our millikelvin-sensitive line survey of TMC-1 has revealed hundreds of unknown features
whose identification challenges our state-of-the-art understanding of the chemistry
of this object \citep{Cernicharo2021b,Cernicharo2021c}.
Line identification in this work was done using the catalogues
MADEX \citep{Cernicharo2012}, CDMS \citep{Muller2005}, and JPL \citep{Pickett1998}.

\begin{figure*}[]
\centering
\includegraphics[scale=0.55,angle=0]{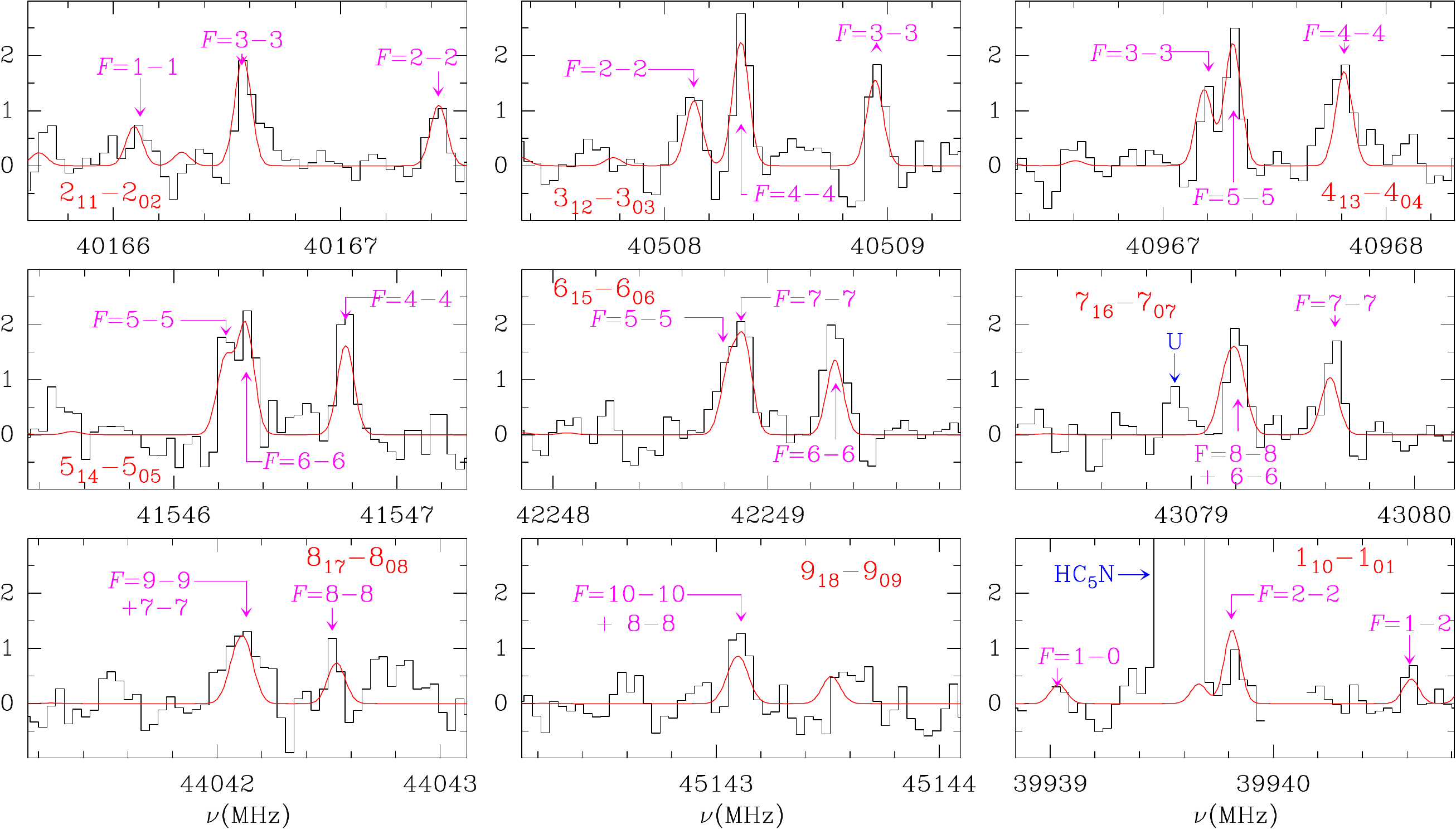}
\caption{Same as Fig. \ref{fig_type_a} but for the  $b$-type transitions of HCSCN.
The red lines show the computed synthetic spectrum for this species for T$_r$=5 K and
a column density of 1.3$\times$\doce.}
\label{fig_type_b}
\end{figure*}

One surprising result in our data is the presence of a few strong unknown features with
intensities around 10 mK. Three of them are in an excellent harmonic relation of 6:7:8. A fit
to the standard Hamiltonian of a linear molecule provides a rotational constant, $B$, of 3082.8334$\pm$0.0029 MHz
and a distortion constant, $D$, of 40.890$\pm$0.027 kHz. Such a large distortion constant points towards
an asymmetric rotor. Hence, these three lines could correspond to the transitions $K_a$=0 of
$J_{up}$=6, 7, and 8 of a molecule with $(B+C)/2$=3082.8 MHz. This suggests that the molecule could
have a total of either four or five N, C, and/or O atoms. 
We explored around these features, looking for
the $K_a$=1 lines. After some iterative work, we found six of these lines that can be fitted with a
reduced number of rotational and distortion constants for a total of nine transitions. These lines are
shown in Fig. \ref{fig_type_a}. The assigned quantum numbers correspond to $a$-type transitions.
The carrier is definitely an asymmetric rotor. Hence, we could expect to have $b$-type transitions if
the molecule has a large $\mu_b$.
From the rotational and distortion constants derived from the $a$-type transitions, we predicted
the spectrum for $b$-type transitions and found nine of them, as shown in Fig. \ref{fig_type_b}.
Moreover,
these $b$-type transitions show a series of lines around the predicted frequencies, which is reminiscent of the hyperfine structure of an
N-nucleus. Assuming this is the case, we fitted all these lines and obtained the rotational
constants provided in Table \ref{rot_constants}. The derived values for $\chi_{aa}$ and $\chi_{bb}$
definitively correspond to a CN group. The hyperfine structure of the $a$-type transitions is not resolved
with our spectral resolution. Hence, the three strongest components appear as a
single feature. No $c$-type transitions were found at the level of sensitivity of our survey.

\begin{table*}
\tiny
\caption{Spectroscopic parameters for HCSCN, HC$^{34}$SCN, and HCSCCH.}
\label{rot_constants}
\centering
\begin{tabular}{|l|cccc|cc|c|}
\hline
Constant            & TMC-1$^a$        & Laboratory $^b$  & Merged fit$^c$    & Laboratory $^{34}$S$^d$ & Bogey1989$^e$& Zaleski2012$^f$&HCSCCH$^g$ \\
\hline
$A$ (MHz)           &  42910.051(11)   &42910.0474(25)    &  42910.0478(21)   & 42604.3412$^h$ &  43314.0052(34)& 42909.9686(83)& 42652.03185(40)    \\
$B$ (MHz)           &   3195.4016(17)  & 3195.40149(55)   &   3195.40171(26)  & 3116.7454(14)&   3205.67686(24) &  3195.3952(17)&  3109.382386(28)   \\
$C$ (MHz)           &   2970.1303(17)  & 2970.13104(54)   &   2970.13067(28)  & 2900.6315(14)&   2975.32470(23) &  2970.1273(18)&  2894.265480(28)   \\
$D_J$ (kHz)         &      1.2767(93)  &   1.2812(76)     &     1.2792(28)    & 1.228(30)    &      1.26557(17) &     1.218(11) &     1.1298295(90)    \\
$D_{JK}$ (kHz)      &   -104.31(33)    &-104.29(14)       &  -104.33(12)      &-102.09(26)   &    -102.0177(24) &   -105.78(38) &  -104.85202(18)     \\
$d_1$ (kHz)         &     -0.2288(64)  &  -0.2228(96)     &    -0.2277(25)    & -0.247(41)   &     -0.228786(42)&    0.2173(93) &     -0.2110838(23)   \\
$\chi_{aa}$ MHz     &     -3.46(16)    &  -3.5118(20)     &    -3.5117(20)    & -3.4952(33)  &                  &  -3.53(20)    &           \\
$\chi_{bb}$ MHz     &      0.83(12)    &   0.8384(22)     &     0.8385(22)    &  0.828(11)   &                  &   0.86(30)    &           \\
\hline
Number of lines     & 32$^i$           & 53               & 85                & 21           & 152              & 36            &   77      \\
$\sigma$(kHz)       & 11.8             & 1.3              & 8.9               & 1.2          & 28.8             & 22.1          & 7.5       \\
$J_{max}$, $Ka_{max}$& 9, 1            & 4,1              & 9,1               & 3,1          & 58,20            &               & 20,5      \\
$\nu_{max}$ (GHz)   & 50.204           & 25.112           & 50.204            & 18.376       & 279.98           &               & 51.3      \\
\hline
\end{tabular}
\tablefoot{\\
Values between parentheses correspond to the uncertainties of the parameters
in units of the last significant digits.\\
\tablefoottext{a}{Fit to the line frequencies measured in TMC-1.} 
\tablefoottext{b}{Fit to the laboratory line frequencies measured in this work.} 
\tablefoottext{c}{Merged fit using our laboratory measurements and the frequencies measured in TMC-1.
This is the recommended set of constants to predict the frequencies of HCSCN.}
\tablefoottext{d}{Fit to the laboratory line frequencies of HC$^{34}$SCN measured in this work.}
\tablefoottext{e}{Rotational constants for HCSCN from \citet{Bogey1989}:
They probably correspond to a vibrational state of HCSCN but not to the ground state of 
this species (see Appendix \ref{lab}). Additional distortion constants
were obtained by these authors, but they are not relevant for the present work.}
\tablefoottext{f}{Rotational constants for HCSCN from \citet{Zaleski2012}. The $d_1$ constant
corresponds in this case to -$\delta_J$.}
\tablefoottext{g}{Rotational constants for HCSCCH derived from a fit to the data
observed for this species by \citet{Brown1982}, \citet{Cabtree2016}, 
and \citet{Margules2020}. 
The full set of distortion constants are provided by \citet{Margules2020}.}
\tablefoottext{h}{Fixed to the theoretical value.}
\tablefoottext{i}{The $a$-type transitions are unresolved and have been fitted to the centroid of the
hyperfine components.}
}
\end{table*}
\normalsize

\begin{figure*}[]
\centering
\includegraphics[scale=0.55,angle=0]{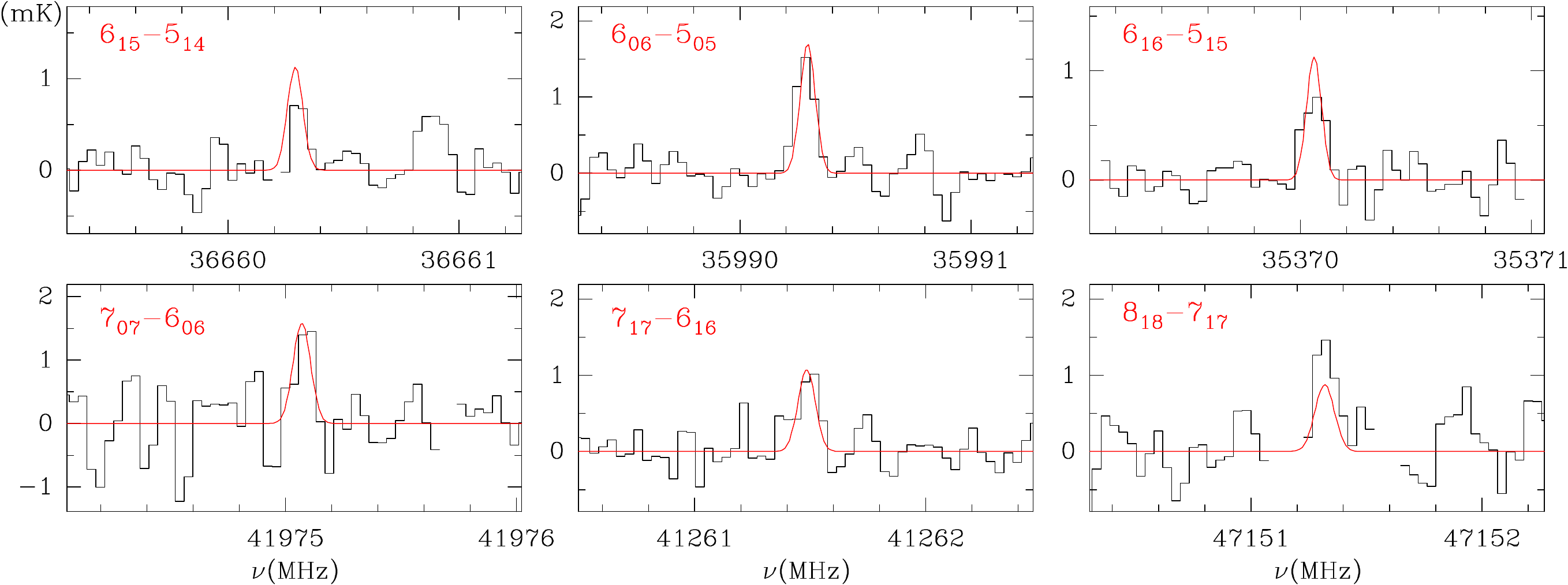}
\caption{Same as Fig. \ref{fig_type_a} but for the observed transitions of HCSCCH towards TMC-1.
The red line shows the computed synthetic spectrum for propynethial assuming $T_r$\,=\,5\,K and
$N$(HCCCHS)\,=\,3.2\,$\times$\,\once.}
\label{fig_hcscch}
\end{figure*}

\subsection{Identifying the carrier: Laboratory microwave spectroscopy experiments}
\label{assig_2st_step}

We explored, using ab initio calculations, a significant number of the combinations of H$_x$C$_y$N$_z$
and H$_x$C$_y$O$_z$ without finding a good agreement with the constants of our new species.
Molecules such as bent-HC$_4$N and $c$-C$_3$HCN have rotational constants close, within 10\%, to ours.
Several years ago we implemented  a species in MADEX that matches our constants rather well. It is HCSCN (see Table \ref{rot_constants}), and it was observed in the
laboratory by \citet{Bogey1989}. Another species with very close
constants to ours (see Table \ref{rot_constants}), HCSCCH,  is also implemented in MADEX,
and we found several of its lines in our data (see Sect. \ref{hcscch}). Hence, HCSCN could
be a good candidate for TMC-1. Unfortunately, with the rotational constants provided by
\citet{Bogey1989}, the species is not detected. We explored the literature for
additional laboratory data for HCSCN and found a contribution to the Ohio State 67th
International Symposium on Molecular Spectroscopy \citep{Zaleski2012}. In the presentation,
the reported rotational constants for HCSCN agree perfectly with those derived in TMC-1
(see Table \ref{rot_constants}); \citet{Zaleski2012} also claim a difference between their data and those
of \citet{Bogey1989}.
The detection of HCSCCH, and the good agreement between the TMC-1 constants and those of
\citet{Zaleski2012}, points to an uncorrect assignment of the HCSCN lines in
the \citet{Bogey1989} work.
A detailed search for a publication of Zaleski et al. data in a refereed journal has, regrettably, been unsuccessful.

In order to confirm that the carrier of our prominent unknown features in TMC-1 is indeed HCSCN,
we performed laboratory measurements. The rotational spectrum of HCSCN was measured by a Balle-Flygare-type FT microwave spectrometer
combined with a pulsed discharge nozzle \citep{Endo1997,Cabezas2016}. HCSCN was produced in a supersonic 
expansion by
a pulsed electric discharge of a gas mixture of CH$_3$CN (0.2\%) and CS$_2$ (0.2\%) diluted in Ar and by applying a 
voltage
of 2.0 kV through the throat of the nozzle source. We observed a total of nine $a$-type and two $b$-type 
transitions (see Fig. \ref{fig_ftmw}) at the predicted frequencies using the TMC-1 observed lines. To confirm the 
identity of the spectral carrier, we observed the HC$^{34}$SCN isotopic species in their natural abundance. Their rotational transitions 
(see Fig. \ref{fig_ftmw}) appear exactly at the expected frequencies considering the isotopic shift for $^{34}$S. All the 
observed lines for HCSCN and HC$^{34}$SCN are shown in Tables \ref{tab_lab_hcscn} and \ref{tab_lab_34S}.

We performed ab initio calculations 
at the MP2/6-311++G(d,p) level of theory
to optimise the HCSCN molecular structure and calculate the electric 
dipole moment components. We obtained $\mu_a$ and $\mu_b$ values of 2.03 and 1.69 D. Details 
of these calculations are given 
in Appendix \ref{lab}. Our calculations also indicate that the observed rotational transitions reported by 
\citet{Bogey1989} corresponds to the lowest excited vibrational state of HCSCN, $\nu_{9}$, 
which lies at 198.1 cm$^{-1}$ above the ground vibrational state.

The merged fit to the TMC-1 frequencies and the fit to our laboratory measurements provide a
set of rotational constants (see Table \ref{rot_constants})
that can be used to predict the frequencies of HCSCN up to
100 GHz with calculated uncertainties below 50 kHz
for $K_a$=0 and 1. Within the present sensitivity of our 
survey, no lines of HC$^{34}$SCN, which are expected to be 20 times weaker than those of HCSCN, 
were found.

\subsection{Column density of HCSCN}

We used a model fitting technique (see \citealt{Cernicharo2021a}) in which the rotational temperature,
the column density, and the $\mu_a$/$\mu_b$ ratio are varied.
A uniform brightness temperature source with a diameter of 80$''$ was assumed \citep{Fosse2001}.
The best fit provides a rotational temperature of
5.0$\pm$0.5\,K, a dipole moment component ratio, $\mu_a$/$\mu_b$, of 0.70$\pm$0.05, and
a column density of (6.2$\pm$0.5)/$\mu_a^2$$\times$\doce, where $\mu_a$ is the component of
the dipole moment along the $a$-axis of the molecule. Adopting the value of the total dipole
moment derived by our ab initio calculations and the $\mu_a$/$\mu_b$ ratio of 0.7 obtained
from our observations, we derive $\mu_a$=2.16\,D. Hence, the column density of HCSCN is
(1.3$\pm$0.1)$\times$\doce. Adopting a column density of H$_2$ of 10$^{22}$ cm$^{-2}$ for
TMC-1 \citep{Cernicharo1987}, the abundance of HCSCN relative to H$_2$ is 1.3\,$\times$\,10$^{-10}$.

The column density of cyano formaldehyde, HCOCN, in TMC-1 is (3.5$\pm$0.5)$\times$\once
(see Appendix \ref{hcocn}). Hence, the
abundance ratio HCSCN/HCOCN is $\sim$4, which is surprising considering that 
the elemental gas-phase abundance of oxygen is expected to be much larger than that of sulphur. 
Moreover, the column density of formaldehyde is 5$\times$10$^{14}$ cm$^{-2}$ \citep{Agundez2013}, while
that of H$_2$CS is 4.7$\times$10$^{13}$ cm$^{-2}$ \citep{Cernicharo2021a}. Hence, 
the abundance ratio HCSCN/HCOCN is a factor of 40 larger
than that of H$_2$CS/H$_2$CO ($\sim$0.1).

\section{Detection of propynethial, HCSCCH} \label{hcscch}
The detection of HCSCN suggests that other derivatives of thioformaldehyde
could be present in TMC-1. The obvious candidate is ethynyl thioformaldehyde (HCSCCH), also known as propynethial. 
This molecule has been observed in the laboratory up to 
630 GHz and $J_{max}$=107 \citep{Brown1982,Cabtree2016,Margules2020}.
Its dipole moment components have been measured to be
$\mu_a$=1.763\,D and $\mu_b$=0.661\,D \citep{Brown1982}. Six lines of this molecule were
found in our data, and they are shown in Fig. \ref{fig_hcscch}. The lines are considerably weaker
than those of HCSCN. We searched
for more transitions of this species, but their expected intensity is below the present sensitivity
of the survey. Nevertheless, the detection of all the strongest lines of \mbox{HCSCCH} in the Q-band is solid
and robust. None of the detected lines is blended with any other feature from known species.
This species was searched for towards different sources, including TMC-1, by \citet{Margules2020} with negative
results. However, the transition they searched for towards TMC-1 has an upper energy level of 13.7\,K and is predicted
to be very weak by our models.
Assuming the
same source parameters and rotational temperature as for HCSCN, we derive a column density
for this species of (3.2$\pm$0.4)$\times$\once, which is eight times lower than the upper 
limit derived by \citet{Margules2020}. 
Therefore, we obtain an abundance ratio HCSCN/HCSCCH of 4$\pm$1. The column density of HCOCCH in this source is (1.5$\pm$0.2)$\times$\doce
(see Appendix \ref{propynal}). Hence, the abundance ratio HCOCCH/HCSCCH is 5$\pm$1, 
which is of the order of the H$_2$CO/H$_2$CS abundance ratio ($\sim$10; see above).

HCSCCH is an isomer of H$_2$CCCS, a molecule recently found in TMC-1 by \citet{Cernicharo2021b}. The
column density of the latter has been derived to be (3.7$\pm$0.3)$\times$\once. Hence, the abundance ratio between both
isomers is $\sim$1. The energy difference between them is 2 kcal mol$^{-1}$, the
H$_2$CCCS isomer being the more stable one \citep{Cabtree2016}.
Finally, we could expect HC$_4$CHS, which could result from the reaction of
C$_4$H with H$_2$CS, to be present in TMC-1. Microwave
laboratory data for this species are available \citep{McCarthy2017}. We searched for its
strongest transitions, but we only derive a 3$\sigma$ upper limit to its column density of
\doce.

\vspace{-0.1cm}
\section{Discussion} \label{discussion}

\begin{figure}
\centering
\includegraphics[width=0.9\columnwidth,angle=0]{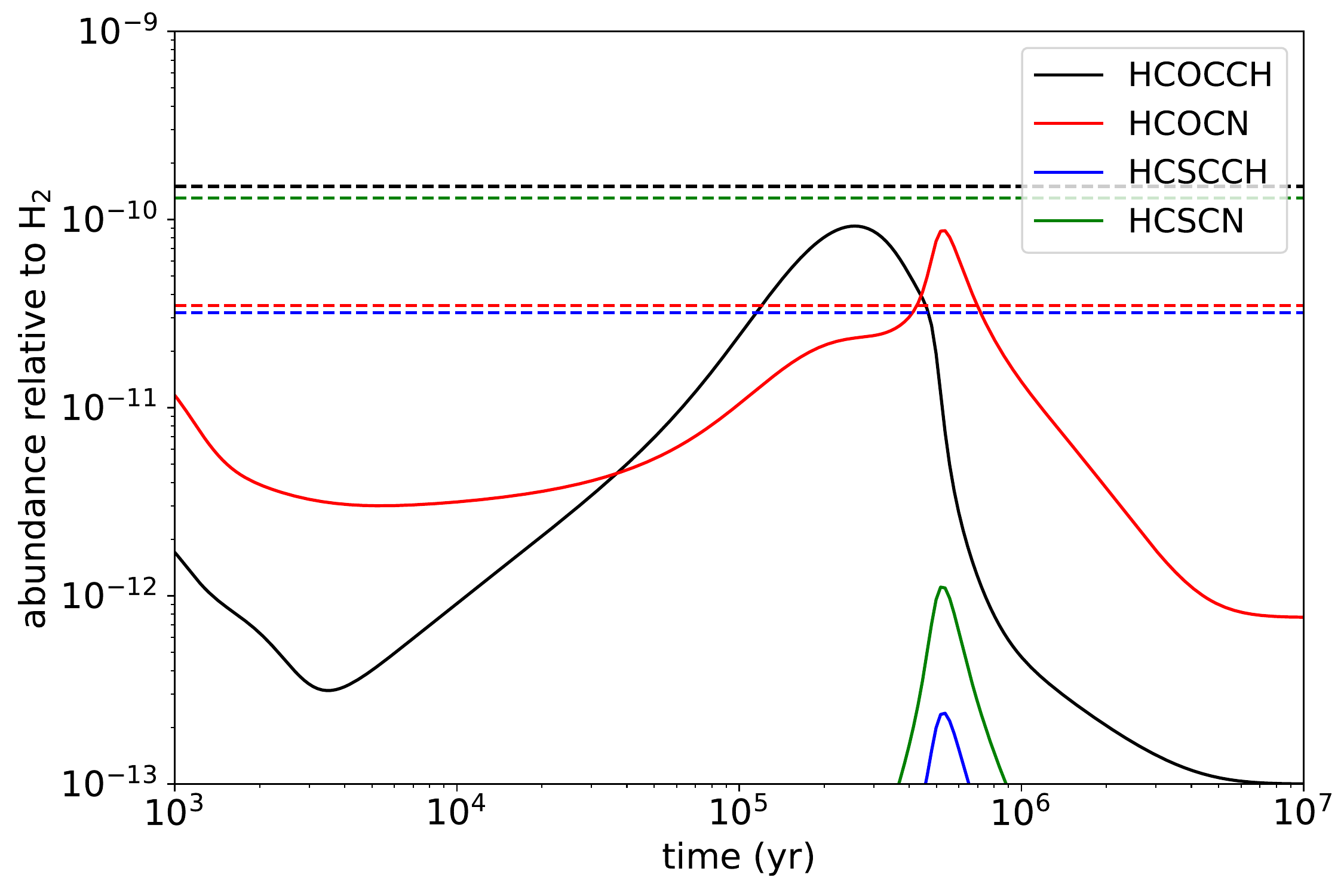}
\caption{Calculated fractional abundances as a function of time. Observed abundances in TMC-1 are indicated by dashed horizontal lines.}
\label{fig:abun}
\end{figure}
The molecules HCSCN and HCSCCH are chemically related as they share an HCS subunit in which the carbon atom is bonded to a CN or a CCH group. Moreover, the oxygen analogues HCOCN and HCOCCH are also chemically related to the two aforementioned molecules since they result from the substitution of the sulphur atom with an oxygen atom. One 
can therefore conceive of similar chemical routes to these four molecules.
A first formation pathway can be provided by
\begin{subequations} \label{reac:o+ch2cch}
\begin{align}
\rm O + CH_2CCH & \rightarrow \rm HCOCCH + H, \label{reac:1a} \\
\rm O + CH_2CN   & \rightarrow \rm HCOCN + H, \label{reac:1b} \\
\rm S + CH_2CCH & \rightarrow \rm HCSCCH + H, \label{reac:1c} \\
\rm S + CH_2CN   & \rightarrow \rm HCSCN + H. \label{reac:1d}
\end{align}
\end{subequations}
These reactions have not been studied, but one could reasonably expect them to be fast at low temperatures. \cite{Loison2016} assume that reaction~(\ref{reac:1a}) is rapid and find that it is the main formation route to HCOCCH in cold interstellar clouds. A second formation route can be provided by
\begin{subequations} \label{reac:c2h+h2co}
\begin{align}
\rm H_2CO + CCH & \rightarrow \rm HCOCCH + H, \label{reac:2a} \\
\rm H_2CO + CN & \rightarrow \rm HCOCN + H, \label{reac:2b} \\
\rm H_2CS + CCH & \rightarrow \rm HCSCCH + H, \label{reac:2c} \\
\rm H_2CS + CN & \rightarrow \rm HCSCN + H. \label{reac:2d}
\end{align}
\end{subequations}
Quantum chemical calculations indicate that reaction~(\ref{reac:2a}) could be 
rapid at low temperatures since the small calculated barrier is within the 
uncertainty of the calculations \citep{Dong2005}, while reaction~(\ref{reac:2b}) 
is barrier-less \citep{Tonolo2020}. These authors also suggest that the reaction 
CN + CH$_3$CHO can plausibly form HCOCN, although it is 
probably less efficient than reaction~(\ref{reac:2b}) since CH$_3$CHO is about 
80 times less abundant than H$_2$CO in TMC-1 \citep{Agundez2013}.

We implemented the aforementioned neutral-neutral reactions with a rate coefficient 
of 10$^{-10}$ cm$^{3}$ s$^{-1}$ in a gas-phase chemical model, in which we adopted 
typical physical conditions of cold dark clouds (see e.g. \citealt{Agundez2013}) 
and used the chemical network {\small UMIST RATE12} \citep{McElroy2013}, revised 
and expanded according to \cite{Loison2017} and \cite{Vidal2017}. The chemical model 
reproduces the observed abundances of the O-bearing molecules HCOCCH 
and HCOCN reasonably well. Moreover, the observed abundance ratio HCOCCH/HCOCN of $\sim$\,4 is of the 
same order as the observed abundance ratios CCH/CN ($\sim$\,10; \citealt{Pratap1997}) 
and CH$_2$CCH/CH$_2$CN ($\sim$\,6; \citealt{Agundez2021}; \citealt{Cabezas2021}), which 
is consistent with the chemical routes proposed. In the case of the S-bearing molecules 
HCSCCH and HCSCN, the chemical model underestimates their abundances by two orders of 
magnitude, which may be related to an excessively low abundance of atomic sulphur calculated by 
the chemical model or to missing formation reactions, such as ion-neutral reactions.

\section{Conclusions}

We have reported the discovery of HCSCN and HCSCCH in TMC-1. While HCSCCH is 
five times less abundant than its oxygen analogue, propynal, we find that 
HCSCN is four times more abundant than HCOCN. Further studies on plausible 
formation reactions of HCSCN and HCSCCH are needed.

\begin{acknowledgements}

This research has been funded by ERC through grant ERC-2013-Syg-610256-NANOCOSMOS. Authors also thank
Ministerio de Ciencia e Innovaci\'on (Spain) for funding support through projects AYA2016-75066-C2-1-P,
PID2019-106235GB-I00 and PID2019-107115GB-C21 / AEI / 10.13039/501100011033. MA thanks Ministerio de
Ciencia e Innovaci\'on for grant RyC-2014-16277.
YE thanks Ministry of Science and Technology, Taiwan, for grant 
No. MOST 108-2113-M-009-025 and MOST 108-2811-M-009-520.

\end{acknowledgements}

\begin{appendix}
\section{Line parameters of the observed transitions of HCSCN and HCSCCH}
\label{line_parameters}
Line parameters for the different molecules studied in this work were obtained by fitting a Gaussian line
profile to the observed data. A window of $\pm$ 15 \kms\, around the local standard of rest velocity of the source was
considered for each transition. The derived line parameters for the two  
molecular species discovered in
this work are given in Table \ref{obs_line_parameters}.
\begin{table}
\tiny
\caption{Line parameters for the lines of HCSCN and HCSCCH.}
\resizebox{0.49\textwidth}{!}{
\label{obs_line_parameters}
\centering
\begin{tabular}{lcrccc}
\hline
\hline
$J\,K_a\,K_c$  &  $\nu_{obs}~^a$&   $\delta_\nu$~$^b$   &$\int T_A^* dv$~$^c$ & $\Delta v$~$^d$ & $T_A^*$~$^e$\\
           & MHz           & kHz         &(mK\,km\,s$^{-1}$)   & (km\,s$^{-1}$)  & (mK)     \\
\hline
HCSCN           &             &             &        &          \\
$5_{14}-4_{13}$ &31386.932(10)& -6& 3.46(25)&1.10(09)& 2.96(27) \\
$6_{16}-5_{15}$ &36310.152(10)&  0& 4.14(17)&0.86(04)& 4.51(22) \\
$6_{06}-5_{05}$ &36958.682(10)& -2& 7.70(20)&0.75(02)& 9.63(27) \\
$6_{15}-5_{14}$ &37661.251(10)&  7& 5.28(25)&0.92(05)& 5.37(30) \\
$7_{17}-6_{16}$ &42357.921(10)& -8& 4.41(26)&0.77(05)& 5.37(31) \\ 
$7_{07}-6_{06}$ &43103.558(10)&  5& 7.77(25)&0.69(03)&10.61(31) \\
$7_{16}-6_{15}$ &43933.874(10)&  0& 4.79(24)&0.69(04)& 6.50(28) \\
$8_{18}-7_{17}$ &48403.924(10)& 10& 2.94(33)&0.58(07)& 4.73(56) \\
$8_{08}-7_{07}$ &49241.600(10)&  8& 6.61(38)&0.68(04)& 9.06(64) \\
$8_{17}-7_{16}$ &50204.498(10)&-14& 3.99(19)&0.40(13)& 9.3(2.7) \\
$2_{11}-2_{02}$ \\
$F=1-1$         &40166.115(20)& 23& 0.42(13)&0.63(22)& 0.63(27)\\
$F=3-3$         &40166.591(20)& 21& 1.34(14)&0.71(11)& 1.78(27)\\
$F=2-2$         &40167.426(20)& -5& 0.82(12)&0.67(11)& 1.14(27)\\
$3_{12}-3_{03}$ \\
$F=2-2$         &40508.116(20)&-16& 1.29(14)&0.83(09)& 1.56(27)\\ 
$F=4-4$         &40508.352(20)& 10& 1.89(12)&0.58(05)& 3.05(27)\\  
$F=3-3$         &40508.954(20)& 10& 1.82(13)&0.67(05)& 2.55(27)\\
$4_{13}-4_{04}$ \\
$F=3-3$         &40967.200(20)& 16& 0.85(14)&0.48(09)& 1.68(28) \\
$F=5-5$         &40967.313(20)&  0& 1.67(17)&0.57(07)& 2.75(28) \\
$F=4-4$         &40967.798(20)&-13& 1.30(17)&0.86(13)& 1.41(28)\\
$5_{14}-5_{05}$ \\
$F=4-4$         &41546.232(20)&  2& 0.96(14)&0.50(15)& 1.79(31)\\
$F=6-6$         &41546.328(20)&  6& 1.29(17)&0.55(08)& 2.20(31)\\
$F=5-5$         &41546.769(20)& -1& 1.32(17)&0.52(08)& 2.40(31)\\
$6_{15}-6_{06}$ \\
$F=5-5$         &42248.836(20)& 17& 0.54(17)&0.50(20)& 1.01(29)\\
$F=7-7$         &42248.879(20)&-11& 1.48(24)&0.68(20)& 2.04(29) \\
$F=6-6$         &42249.309(20)& -5& 1.83(16)&0.83(17)& 2.08(29) \\
$7_{16}-7_{07}$ &\\
$F=8-8 + F=6-6$ &43079.209(20)& 25& 1.22(12)&0.58(07)& 1.94(27) \\
$F=7-7$         &43079.628(20)&  6& 1.08(14)&0.55(09)& 1.83(27) \\
$8_{17}-8_{08}$ \\
$F= 9-9 + 7-7$  &44042.119(20)&  9& 1.17(35)&1.05(30)& 1.05(35) \\
$9_{18}-9_{09}$ \\
$F=10-10 + 8-8$ &45143.106(20)& 10& 1.04(20)&0.62(13)& 1.58(34) \\
\hline
HCSCCH\,$^f$  &             &   &         &        &          \\
$6_{16}-5_{15}$ &35370.056(10)&  0& 0.68(12)&0.87(15)& 0.73(17) \\
$6_{06}-5_{05}$ &35990.280(10)&-16& 1.37(16)&0.82(11)& 1.56(16) \\
$6_{15}-5_{14}$ &36660.308(20)& 32& 0.52(10)&0.57(14)& 0.86(18) \\
$6_{25}-5_{24}$ &36021.980(20)&-18& 0.40(14)&0.65(25)& 0.57(20) \\
$7_{07}-6_{06}$ &41975.060(20)& 28& 0.93(33)&0.54(20)& 1.61(35) \\
$7_{17}-6_{16}$ &41261.501(20)& 21& 0.56(17)&0.56(19)& 0.94(24) \\
$8_{18}-7_{17}$ &47151.323(20)& 39& 2.02(67)&0.91(25)& 2.10(44) \\
\hline
\hline
\end{tabular}
}
\tablefoot{\\
Values between parentheses correspond to the uncertainties of the parameters
in units of the least significant digits.\\
\tablefoottext{a}{Observed frequency assuming a v$_{LSR}$ of 5.83 \kms.
The hyperfine splitting for $a$-type transitions is unresolved.}\\
\tablefoottext{b}{Observed minus calculated frequencies using the rotational constant
of the merged fit provided in Table \ref{rot_constants}.}\\
\tablefoottext{c}{Integrated line intensity in mK\,km\,s$^{-1}$.}\\
\tablefoottext{d}{Line width at half intensity derived by fitting a Gaussian function to
the observed line profile (in km\,s$^{-1}$).}\\
\tablefoottext{e}{Antenna temperature in millikelvin.}\\
\tablefoottext{f}{For this species, predicted frequencies are obtained from the
rotational and distortion constants derived from a fit to the laboratory data of
\citet{Brown1982} and \citet{Cabtree2016}.}\\
}
\end{table}  
\normalsize

\section{Laboratory experiments and quantum chemical calculations for HCSCN}
\label{lab}

\begin{figure}
\includegraphics[angle=0,width=\columnwidth]{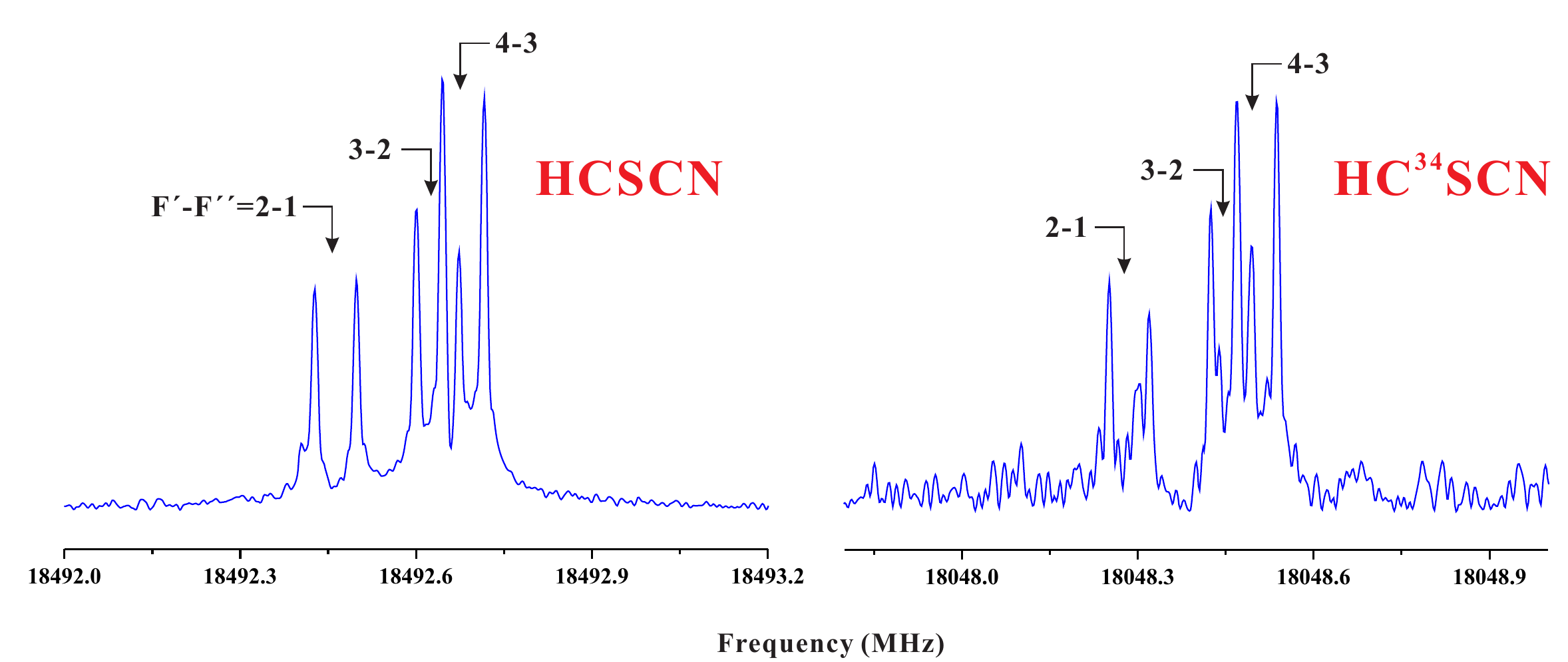}
\caption{\label{fig_ftmw} Fourier transform microwave spectra of HCSCN and HC$^{34}$SCN showing the 3$_{0,3}$-2$_{0,2}$ rotational transition. 
The spectra show the hyperfine components, labelled with the corresponding values of quantum numbers $F'$-$F''$, clearly 
resolved. The coaxial arrangement of the adiabatic expansion and the resonator axis produces an instrumental Doppler doubling. 
The resonance frequencies are calculated as the average of the two Doppler components.}
\end{figure}

The rotational spectrum  of HCSCN was observed in the 12 to 25 GHz frequency region. 
The molecule HCSCN was produced in a supersonic expansion at a rotational temperature of $\thicksim$2K.
Hence, only the rotational transitions with $K_{a}$= 0 and 1 could be observed. Each rotational transition is split 
into several hyperfine components because of the nuclear quadrupole coupling effects produced by the presence 
of a nitrogen nucleus, as can be seen in Fig. \ref{fig_ftmw}. The interaction at the 
nitrogen nucleus of the 
quadrupole moment with the electric field gradient created by the rest of the molecular charges causes 
the coupling of the nuclear spin moment to the overall rotational momentum \citep{Gordy1984}. In total, 
we observed 53 hyperfine components that correspond to 11 pure rotational transitions (see Table \ref{tab_lab_hcscn}). 
Rotational, centrifugal, and nuclear quadrupole coupling constants were determined by fitting the transition frequencies, 
using the SPFIT program \citep{Pickett1991}, to a Watson’s $A$-reduced Hamiltonian for asymmetric top molecules, with 
the following form \citep{Watson1977}: $H$ = $H_R$ + $H_Q$, where $H_R$ contains rotational and centrifugal distortion 
parameters and $H_Q$ the quadrupole coupling interactions. The energy levels involved in each transition are labelled 
with the quantum numbers $J$, $K_{a}$, $K_{c}$, and $F$, where $F$ = $J$ + $I$(N) and $I$(N) = 1. The analysis 
rendered the experimental constants listed in Table \ref{rot_constants}. For the HC$^{34}$SCN isotopic species we observed 
a total of 21 hyperfine components, which were analysed in the same manner as those for HCSCN. The observed frequencies 
are listed in Table \ref{tab_lab_34S} and parameters derived from the fit in Table \ref{rot_constants}.

Structural optimisation calculations were carried out using the M{\o}ller-Plesset post-Hartree-Fock 
method, MP2 \citep{Moller1934}, and the Pople basis set 6-311++G(d,p) \citep{Frisch1984}. In addition, 
we performed anharmonic frequency calculations to estimate the rotational constants of the vibrationally excited 
states of HCSCN. These can be estimated using the first-order vibration-rotation constants $\alpha_i$ that define 
the vibrational dependence of rotational constants $B_{\nu}$= $B_{e}$ - $\sum_i$ $\alpha_i$ ($\nu_i$ + 1/2), 
where $B_{\nu}$ and $B_{e}$ are substitutes for all three rotational constants in a given excited state and in equilibrium, 
respectively, and $\nu_i$ is the vibrational quantum number. Using the rotational constants from the merged fit 
shown in Table \ref{rot_constants} and the $\alpha_i$ constants from our calculations, we derived the rotational 
constants for all the vibrationally excited states of HCSCN. The results are shown in Table \ref{vib_states}. All 
the calculations were performed using the Gaussian 16 program package \citep{Frisch2016}.

As can be seen, the estimated constants for the $\nu_{9}$ are very close to those reported by 
\citet{Bogey1989}. This fact and the low energy of this excited state 
suggest that \citet{Bogey1989} identified this 
state instead of the ground state of HCSCN in their experiment. The experiment conducted by \citet{Bogey1989} was done at room 
temperature and with the HCSCN precursor heated up to 593 K, which can also allow the population of this low energy vibrational state.

\label{ftmw}

\begin{table}
\tiny
\caption{Laboratory observed transition frequencies for HCSCN.}
\label{tab_lab_hcscn}
\centering
\begin{tabular}{cccccccccr}
\hline
\hline
 $J'$ & $K'_a$ & $K'_c$ & $J''$ & $K''_a$ & $K''_c$ &  $F'$ & $F''$ & $\nu_{obs}$  &  Obs-Calc \\
   &   &        &      &        &        &                  &    &   (MHz)      &   (MHz)      \\
\hline
2 & 1 & 2 &  1 & 1 & 1 & 2 & 1 & 12105.301  &   0.002  \\
2 & 1 & 2 &  1 & 1 & 1 & 2 & 2 & 12105.550  &  -0.001  \\
2 & 1 & 2 &  1 & 1 & 1 & 3 & 2 & 12106.412  &   0.001  \\
2 & 1 & 2 &  1 & 1 & 1 & 1 & 1 & 12106.637  &   0.000  \\
2 & 1 & 2 &  1 & 1 & 1 & 1 & 0 & 12107.266  &   0.001  \\
2 & 0 & 2 &  1 & 0 & 1 & 2 & 2 & 12329.013  &   0.000  \\
2 & 0 & 2 &  1 & 0 & 1 & 1 & 0 & 12329.192  &   0.000  \\
2 & 0 & 2 &  1 & 0 & 1 & 2 & 1 & 12330.068  &   0.001  \\
2 & 0 & 2 &  1 & 0 & 1 & 3 & 2 & 12330.145  &   0.001  \\
2 & 0 & 2 &  1 & 0 & 1 & 1 & 1 & 12331.825  &   0.000  \\
2 & 1 & 1 &  1 & 1 & 0 & 2 & 1 & 12555.828  &   0.001  \\
2 & 1 & 1 &  1 & 1 & 0 & 1 & 1 & 12556.245  &  -0.002  \\
2 & 1 & 1 &  1 & 1 & 0 & 2 & 2 & 12556.627  &  -0.002  \\
2 & 1 & 1 &  1 & 1 & 0 & 3 & 2 & 12556.899  &   0.000  \\
2 & 1 & 1 &  1 & 1 & 0 & 1 & 0 & 12558.252  &   0.000  \\
3 & 1 & 3 &  4 & 0 & 4 & 4 & 5 & 14504.659  &  -0.001  \\
3 & 1 & 3 &  4 & 0 & 4 & 3 & 4 & 14504.683  &   0.002  \\
3 & 1 & 3 &  4 & 0 & 4 & 2 & 3 & 14504.763  &  -0.001  \\
3 & 0 & 3 &  2 & 0 & 2 & 3 & 3 & 18491.504  &  -0.001  \\
3 & 0 & 3 &  2 & 0 & 2 & 2 & 1 & 18492.463  &   0.000  \\
3 & 0 & 3 &  2 & 0 & 2 & 3 & 2 & 18492.637  &   0.002  \\
3 & 0 & 3 &  2 & 0 & 2 & 4 & 3 & 18492.681  &   0.001  \\
3 & 0 & 3 &  2 & 0 & 2 & 2 & 2 & 18494.221  &  -0.001  \\
3 & 1 & 3 &  2 & 1 & 2 & 3 & 3 & 18157.527  &  -0.001  \\
3 & 1 & 3 &  2 & 1 & 2 & 3 & 2 & 18158.388  &   0.001  \\
3 & 1 & 3 &  2 & 1 & 2 & 2 & 1 & 18158.648  &  -0.002  \\
3 & 1 & 3 &  2 & 1 & 2 & 4 & 3 & 18158.714  &   0.002  \\
3 & 1 & 3 &  2 & 1 & 2 & 2 & 2 & 18159.986  &   0.000  \\
3 & 1 & 2 &  2 & 1 & 1 & 3 & 3 & 18833.878  &  -0.001  \\
3 & 1 & 2 &  2 & 1 & 1 & 3 & 2 & 18834.150  &   0.001  \\
3 & 1 & 2 &  2 & 1 & 1 & 4 & 3 & 18834.453  &   0.002  \\
3 & 1 & 2 &  2 & 1 & 1 & 2 & 1 & 18834.499  &  -0.003  \\
3 & 1 & 2 &  2 & 1 & 1 & 2 & 2 & 18834.920  &  -0.001  \\
2 & 1 & 2 &  3 & 0 & 3 & 2 & 2 & 20996.955  &   0.001  \\
2 & 1 & 2 &  3 & 0 & 3 & 3 & 4 & 20998.225  &   0.000  \\
2 & 1 & 2 &  3 & 0 & 3 & 1 & 2 & 20998.288  &  -0.003  \\
2 & 1 & 2 &  3 & 0 & 3 & 2 & 3 & 20998.540  &  -0.001  \\
2 & 1 & 2 &  3 & 0 & 3 & 3 & 3 & 20999.402  &   0.002  \\
4 & 1 & 4 &  3 & 1 & 3 & 4 & 4 & 24208.975  &  -0.002  \\
4 & 1 & 4 &  3 & 1 & 3 & 4 & 3 & 24210.163  &   0.001  \\
4 & 1 & 4 &  3 & 1 & 3 & 3 & 2 & 24210.237  &   0.001  \\
4 & 1 & 4 &  3 & 1 & 3 & 5 & 4 & 24210.310  &   0.001  \\
4 & 1 & 4 &  3 & 1 & 3 & 3 & 3 & 24211.834  &  -0.001  \\
4 & 0 & 4 &  3 & 0 & 3 & 4 & 4 & 24651.071  &  -0.001  \\
4 & 0 & 4 &  3 & 0 & 3 & 3 & 2 & 24652.176  &   0.000  \\
4 & 0 & 4 &  3 & 0 & 3 & 4 & 3 & 24652.246  &   0.000  \\
4 & 0 & 4 &  3 & 0 & 3 & 5 & 4 & 24652.279  &   0.002  \\
4 & 0 & 4 &  3 & 0 & 3 & 3 & 3 & 24653.760  &  -0.001  \\
4 & 1 & 3 &  3 & 1 & 2 & 4 & 4 & 25110.542  &   0.001  \\
4 & 1 & 3 &  4 & 3 & 1 & 2 & 3 & 25111.114  &   0.000  \\
4 & 1 & 3 &  3 & 3 & 1 & 2 & 2 & 25111.228  &  -0.001  \\
4 & 1 & 3 &  5 & 3 & 1 & 2 & 4 & 25111.249  &   0.002  \\
4 & 1 & 3 &  3 & 3 & 1 & 2 & 3 & 25112.001  &  -0.002  \\
\hline
\hline
\end{tabular}
\end{table}
\normalsize

\begin{table}
\tiny
\caption{Laboratory observed transition frequencies for HC$^{34}$SCN.}
\label{tab_lab_34S}
\centering
\begin{tabular}{cccccccccr}
\hline
\hline
 $J'$ & $K'_a$ & $K'_c$ & $J''$ & $K''_a$ & $K''_c$ &  $F'$ & $F''$ & $\nu_{obs}$  &  Obs-Calc \\
   &   &        &      &        &        &                  &    &   (MHz)      &   (MHz)      \\
\hline
2 & 1 & 2 & 1 & 1 & 1 & 2 &  1 &  11818.143  & ~  0.001  \\
2 & 1 & 2 & 1 & 1 & 1 & 2 &  2 &  11818.391  & ~  0.000  \\
2 & 1 & 2 & 1 & 1 & 1 & 3 &  2 &  11819.249  & ~  0.001  \\
2 & 0 & 2 & 1 & 0 & 1 & 2 &  2 &  12032.778  & ~ -0.002  \\
2 & 0 & 2 & 1 & 0 & 1 & 1 &  0 &  12032.957  & ~  0.000  \\
2 & 0 & 2 & 1 & 0 & 1 & 2 &  1 &  12033.829  & ~  0.000  \\
2 & 0 & 2 & 1 & 0 & 1 & 3 &  2 &  12033.905  & ~  0.001  \\
2 & 0 & 2 & 1 & 0 & 1 & 1 &  1 &  12035.577  & ~ -0.001  \\
2 & 1 & 1 & 1 & 1 & 0 & 2 &  1 &  12250.356  & ~  0.002  \\
2 & 1 & 1 & 1 & 1 & 0 & 3 &  2 &  12251.421  & ~  0.000  \\
3 & 1 & 3 & 2 & 1 & 2 & 3 &  2 &  17727.695  & ~ -0.001  \\
3 & 1 & 3 & 2 & 1 & 2 & 2 &  1 &  17727.956  & ~  0.000  \\
3 & 1 & 3 & 2 & 1 & 2 & 4 &  3 &  17728.018  & ~  0.000  \\
3 & 0 & 3 & 2 & 0 & 2 & 3 &  3 &  18047.332  & ~ -0.001  \\
3 & 0 & 3 & 2 & 0 & 2 & 2 &  1 &  18048.285  & ~ -0.001  \\
3 & 0 & 3 & 2 & 0 & 2 & 3 &  2 &  18048.460  & ~  0.002  \\
3 & 0 & 3 & 2 & 0 & 2 & 4 &  3 &  18048.503  & ~  0.001  \\
3 & 0 & 3 & 2 & 0 & 2 & 2 &  2 &  18050.037  & ~  0.001  \\
3 & 1 & 2 & 2 & 1 & 1 & 3 &  2 &  18375.983  & ~  0.002  \\
3 & 1 & 2 & 2 & 1 & 1 & 4 &  3 &  18376.284  & ~  0.000  \\
3 & 1 & 2 & 2 & 1 & 1 & 2 &  1 &  18376.331  & ~ -0.003  \\
\hline
\hline
\end{tabular}
\end{table}
\normalsize

\begin{table}
\tiny
\caption{Predicted rotational constants for all the vibrationally excited states of HCSCN$^\#$.}
\label{vib_states}
\centering
\begin{tabular}{crccc}
\hline
\hline
Vib. State & Freq. &  $A$     &   $B$   &  $C$ \\
\hline
$\nu_{1}$ & 3024.3 & 43030.42   & 3189.86        & 2966.02   \\
$\nu_{2}$ & 2121.3 & 42778.69   & 3182.66        & 2958.49   \\
$\nu_{3}$ & 1332.8 & 42588.35   & 3199.20        & 2969.19   \\
$\nu_{4}$ & 1127.2 & 42228.73   & 3197.73        & 2970.08   \\
$\nu_{5}$ &  896.1 & 43771.31   & 3181.79        & 2957.06   \\
$\nu_{6}$ &  827.6 & 42336.24   & 3195.16        & 2972.17   \\
$\nu_{7}$ &  511.9 & 43822.16   & 3192.80        & 2967.49   \\
$\nu_{8}$ &  344.9 & 42726.74   & 3196.93        & 2975.02   \\
$\nu_{9}$ &  198.1 & 43146.20   & 3207.96        & 2976.79   \\
Bogey1989$^a$ &  -  & 43314.00   & 3205.68    & 2975.32   \\
Ground$^b$&     - & 42910.05   & 3195.40    & 2970.13   \\
\hline
\hline
\end{tabular}
\tablefoot{\\
\tablefoottext{\#}{Experimental constants derived in this work and those from 
\citet{Bogey1989} are shown for comparison.}\\
\tablefoottext{a}{Experimental rotational constants from \citet{Bogey1989}.}\\
\tablefoottext{b}{Experimental rotational constants from the merged fit of Table \ref{rot_constants}.}\\
}
\end{table}     
\normalsize

\begin{figure}[]
\centering
\includegraphics[scale=0.6,angle=0]{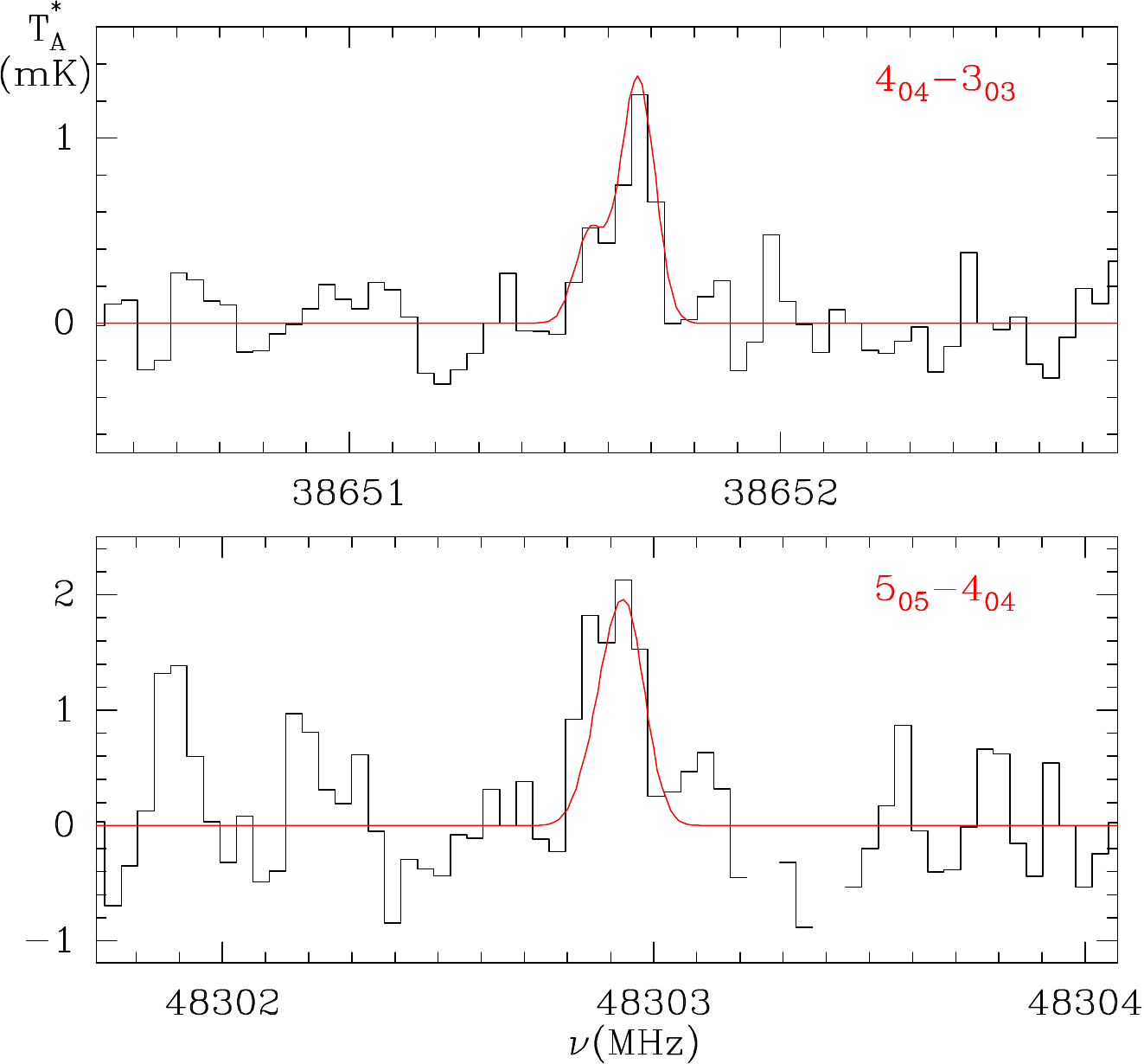}
\caption{Same as Fig. \ref{fig_type_a} but for the observed transitions of HCOCN towards TMC-1.
The red line shows the computed synthetic spectrum for propynethial assuming $T_r$\,=\,5\,K and
$N$(HCOCN)\,=\,3.5\,$\times$\,\once.}
\label{fig_hcocn}
\end{figure}

\section{Cyano formaldehyde}
\label{hcocn}
Cyano formaldehyde, HCOCN, was detected towards Sgr B2(N) by \citet{Remijan2008}.
Laboratory spectroscopy on this molecule was achieved by \citet{Bogey1995}.
A total dipole moment of 2.8\,D was adopted from ab initio calculations
\citep{Csaszar1989}, and the $\mu_a$ and $\mu_b$ components derived from
the ratio $\mu_a$/$\mu_b$=0.63 found by \citet{Bogey1995} were used. In our data
only the $4_{04}-3_{03}$ and $5_{05}-4_{04}$ transitions are detected. They are rather
weak compared to those of HCSCN. Assuming a rotational temperature of 5\,K, as derived
for HCSCN, all the
remaining lines are below our detection limit. Figure \ref{fig_hcocn} shows
the lines and the model obtained for a column density for this species
of (3.5$\pm$0.5)$\times$\once.

\section{Propynal, HCOCCH}
\label{propynal}
\begin{figure}[]
\centering
\includegraphics[scale=0.6,angle=0]{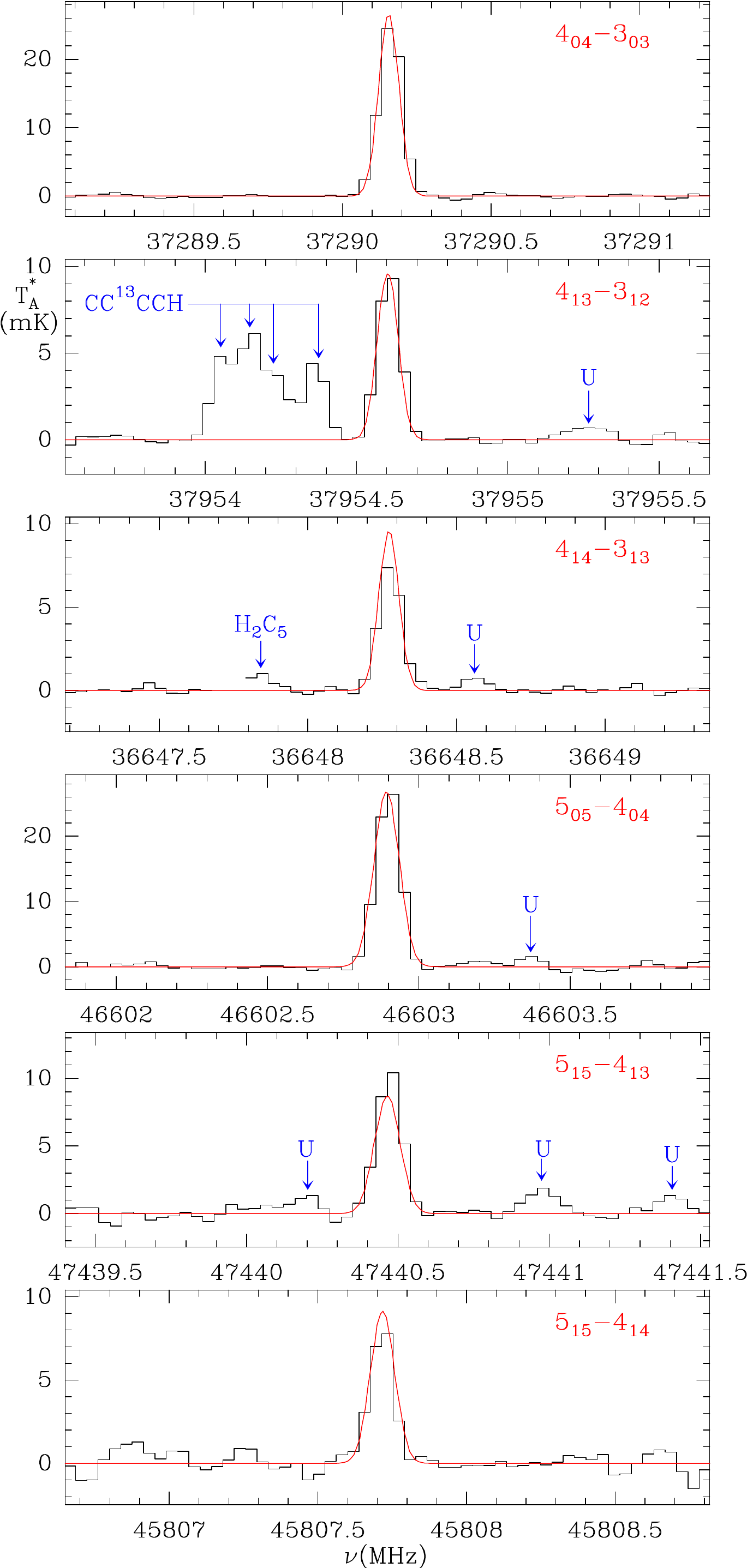}
\caption{Same as Fig. \ref{fig_type_a} but for the observed transitions of HCOCCH towards TMC-1.
The red line shows the computed synthetic spectrum for propynal assuming
$T_r$\,=\,5\,K for the $K_a$=0 lines and $T_r$=4\,K for the $K_a$=1 lines. The resulting
column density is $N$(HCOCCH)\,=\,1.5\,$\times$\,\doce.}
\label{fig_hcccho}
\end{figure}

A rich literature of laboratory work  exists for this species. It has been
summarised by \citet{Jabri2020}. The molecule was detected in TMC-1
by \citet{Irvine1988} and towards several cold interstellar clouds by
\citet{Loison2016}. In TMC-1 the lines are very strong. They are shown
in Fig. \ref{fig_hcccho}. A fit to all the observed lines provides a rotational temperature
of 4.5$\pm$0.5\,K and a column density for propynal of (1.5$\pm$0.3)$\times$\,\doce, in
excellent agreement with the value obtained by \citet{Irvine1988}. It is, however,
a factor of two larger than that obtained by \citet{Loison2016}. This difference
is probably due to the value they adopted for the rotational temperature
of the observed lines, $T_r$=10\,K, and the fact that the
energies of the upper levels of these transitions are larger than 10\,K. Using their
integrated line intensities and a rotational temperature of 5\,K, their column density
would be practically identical to ours. We would like to note that in our fit
shown in Fig. \ref{fig_hcccho} the synthetic spectrum for the $K_a$=0 lines has been
computed for a $T_r$ of 5\,K. However, for the $K_a$=1 lines it has been calculated
with a rotational temperature of 4\,K. This difference is within the uncertainty of
the derived rotational temperature.

\end{appendix}

\end{document}